\documentclass[12pt]{iopart}
\usepackage{iopams,setstack,epsfig}
\include{psfig}
\def\egal{\overset{\mathrm{def}}{=}}
\def\dfrac#1#2{\displaystyle\frac{#1}{#2}}
\def\binom#1#2{\left(#1\atop #2\right)}

\begin{document}

\title{Small-tau expansion for the form factor of glued quantum star graphs.}
\author{Marie-Line Chabanol}
\address{Institut de Math\'ematiques de Bordeaux\\UMR 5251 CNRS-Universit\'e Bordeaux1\\France}
\ead{Marie-Line.Chabanol@math.u-bordeaux1.fr}

\begin{abstract}{We compute the small-tau expansion up to the third order
for the form factor of two glued quantum star graphs with Neumann boundary 
conditions, by taking into account only the most backscattering orbits.
 We thus show that the
glueing has no effect if the number of glueing edges is negligible compared to the number of edges of the graph, whereas it has an effect on the $\tau^2$ term when the numbers of glueing and non glueing edges are of the same order.
}
\end{abstract}
\submitto{\JPA}

\section{Introduction}

The study of the Laplacian on a metric graph, a concept known as
{\it quantum graphs}, now serves as a toy model for quantum
chaos \cite{KS1, GS,BG}. Indeed, there exists an exact trace formula
relating eigenvalues and periodic orbits. Moreover, depending on the
graphs, exact computations of these orbits may be possible, whereas
they are out of reach in most dynamical systems. It has thus been
shown \cite{KS1,KS2,GA} that spectral statistics of simple generic graphs
follow random matrix statistics when the size of the graph tends to
infinity, as expected for chaotic quantum systems. $V$-star graphs
with Neumann boundary conditions (graphs
formed by a central vertex connected to $V$ other vertices by edges of
different lengths) play a
special role because of the high degeneracy of their periodic
orbits. As could be expected, this degeneracy breaks the random matrix
statistics  : this has been shown by  the computation of the two-point
correlation function (\cite{ThB,Ber}). We will investigate here what
happens when two star graphs are glued together.
We will first show that glueing two star graphs with $o(V)$
incommensurate bonds has no effect on the form factor when $V$ tends
to infinity, and we will
next compute the small $\tau$ expansion of two $V$-star graphs glued
by $O(V)$  bonds. 

Spectral statistics for quantum graphs and the trace formula relating
them to periodic orbits will be presented in the first part, as well as
a presentation of the model we will be dealing with. The second
part will recall the small $\tau$ expansion of the
form factor for star graphs as obtained in
\cite{Ber}. In a third part,
we will explain our computations in the glued case.

\section{Quantum graphs : the trace formula, and the quasar model}

We will start by some vocabulary and notations.
Let $G=(E,V)$ be a graph with a metric structure : to each edge 
$(i,j)\in E\subset V\times V$
is assigned a length
$l_{ij}$, such that $l_{ij} = l_{ji}$; although the graph is supposed
to be non oriented,
that is $(i,j)\in E \Rightarrow (j,i)\in E$, and $l_{ij} = l_{ji}$, we
will consider the edges to be oriented : $(i,j)$ is different form
$(j,i)$,
 it really describes the edge going from $i$
to $j$). On each edge $(i,j)$, one can thus define a
coordinate $x$ such that $x=0$ corresponds to the vertex $i$, and
$x=l_{ij}$ corresponds to the vertex $j$. A {\em periodic orbit of period
$n$} is a set of $n$ edges $(p_1,\ldots,p_n)$ such that $p_i$ ends where
$p_{i+1}$ starts (as well as $p_n$ and
$p_1$). A periodic orbit is  called {\em primitive} if it is not the
repetition of a shorter periodic orbit. A primitive orbit repeated $r$
times is a non-primitive orbit with {\em repetition number
  $r$}. $\{\mathcal{l}\}$ will denote the equivalence class of all
orbits of length $l$.
On each edge $(i,j)$, one is looking for the spectrum of  the 
Laplacian. In other words, one wants to find $\lambda$ and $\psi_{ij}$
such that
$-\frac{\rmd^2\psi_{ij} }{\rmd x^2} = \lambda^2 \psi_{ij}(x)$. As one looks
for eigenfunctions defined on the whole graph, one imposes continuity
relations at each vertex, $\psi_{ij}(0) = \psi_{ik}(0)$. Moreover, the
function should have a unique value on a given point, regardless of
the sense of the edge it belongs to : hence one
wants $\psi_{ij}(x) = \psi_{ji}(l_{ij}-x)$. This actually corresponds
to the Neumann condition on each vertex $\sum_{j}
\frac{d\psi_{ij}}{dx} = 0$. It is then a simple exercise to check that the
eigenvalues $\lambda$ are the solutions of $\det(I-e^{-i\lambda L}
S) =0$, where $S$ and $L$ are $|E|\times|E|$ matrices : $L$ is
diagonal with the
length of each edge as diagonal element, and $S$ is defined by
$S_{(i,j),(j,k)} = -\delta_{i,k} + \frac{2}{v_j}$, where $v_j$ is the
number of $j$'s neighbours.

The trace formula as derived in \cite{KS1} states that if $d(\lambda) =
\sum_n \delta(\lambda - \lambda_n)$ is the spectral density, then
$d(\lambda) = \frac{L}{2\pi} + \frac{1}{\pi} \sum_{n} \sum_{p \in P_n}
\frac{l_p}{r_p} A_p \cos (\lambda l_p)$.
Here $L$ is the total length of all edges, $P_n$ is the set of all
periodic orbits of period $n$ up to cyclic reordering (that is
$p_0,p_1,p_2$ and $p_1,p_2,p_0$ are the same orbits), $l_p$ is the
length of the orbit, $r_p$ its repetition number, and $A_p =
\prod_{i=1}^n S_{p_{i},p_{i+1}}$.

A $V$-star graph is a  graph with vertices $\{0,..V\}$ and
edges $E=\{(0,i),(i,0), 1\leq i \leq V\}$ : the $V$ vertices are all connected
to the center $0$. 
 The $S$-matrix elements for such a graph are
$S_{(0,i),(i,0)} = 1$ (this corresponds to trivial scattering),
$S_{(i,0),(0,i)} = -1+\frac{2}{V}$ (backscattering) and
$S_{(i,0),(0,j)} = \frac{2}{V}$ (normal scattering). 
The lengths will be taken so that they are incommensurate, and that
their distribution is peaked around 1 : for instance, they can be
chosen randomly, uniformly in
$[1-\frac{1}{2V},1+\frac{1}{2V}]$, each length being independent from
the others. 

We will consider two such $V_1$-star graphs and $V_2$-star graphs, and
connect them with $M$
edges linking the two centers. We will call the resulting graph a $(V_1,V_2,M)$ quasar graph,
and denote $V=V_1+V_2+M$. The
edges of the two star graphs will be denoted by roman letters ($a,b,$
and so on) and the glueing edges by greek letters ($\epsilon, \zeta,$
and so on). 

\begin{figure}
\centering\epsfig{file=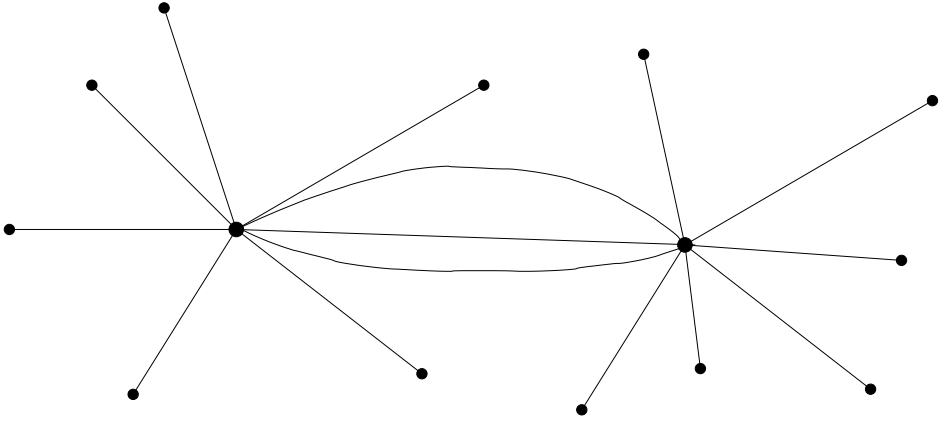,height=5.5cm}
\caption{A quasar graph with $V_1=V_2=6$ and $M=3$}
\label{quas} 
\end{figure}
The $M$ glueing bonds are also supposed to have incommensurate lengths, chosen
randomly in $[1-\frac{1}{2V},1+\frac{1}{2V}]$. Hence a periodic orbit
of period $2n$  has  a length in $[2n-\frac{n}{V},2n+\frac{n}{V}]$; such intervals for different $n$ less than $V$ do not overlap. 
Of course, in a periodic orbit of period $2n$, each non glueing edge is
visited an even number of times (and there is are trivial backscatterings),
whereas a glueing edge can be visited an odd number of times 
(but of course the total number
of visits oof glueing edges has to be even).  The contribution $A_p$ of such 
an 
orbit will depend on the number of backscatterings and normal scatterings, and 
will also depend on the center on which the scattering takes place : it will  be a product of factors $r_i = (-1+\frac{2}{V_i+M})$ (for backscatterings on graph $i$) and $t_i = \frac{2}{V_i+M}$ (for normal scatterings on graph $i$). 

The interesting limit will be the limit $V,n$ tends to infinity with $\tau =\frac{n}{V} $ fixed. In this limit, orbits with the biggest contribution $A_p$ will be orbits
with the largest number of backscatterings. We will
suppose that $V_1 = \nu_1 V$ and $V_2 = \nu_2 V$, and consider two cases : $M =
\nu_3 V$, or $M=o(V)$.


\section{Form factor for the $V$-star graph}

The 2-point correlation function is defined as $R_2(x) =\left(\frac{2\pi}{L}\right)^2 \langle d(\lambda) d(\lambda - \frac{2\pi
  x}{L})\rangle$. The brackets denote a mean value with respect  
to the $\lambda$'s, that is $\langle f\rangle
=\lim_{\Lambda \rightarrow \infty} \frac{1}{2\Lambda}
\int_{-\Lambda}^{\Lambda} {f(\lambda)} d\lambda$. 
Using the trace formula and performing the integral, one gets
\begin{eqnarray*}
R_2(x) =  1+\frac{2}{L^2} \sum_{p,p'}\frac{l_p l_{p'}}{r_p
r_{p'}} A_p A_{p'}  \delta_{l_p - l_{p'}}
\cos(\frac{2\pi x l_p}{L}),
\end{eqnarray*}
where the sum is over the pairs of periodic orbits $(p,p')$, up to cyclic
permutations, of  lengths $l_p$ and $l_{p'}$ and
repetition numbers $r_p$ and $r_{p'}$.

The form factor is defined by $K_{st}(\tau) = \int_{-\infty}^{+\infty}
(R_2(x)-1) 
\exp(2i\pi x \tau)
\rmd\tau$. When $V$ is large,  $K^* (\tau)$ for $\tau \in [\frac{n}{V}
-\frac{1}{2V}, \frac{n}{V} + \frac{1}{2V}]$ is well approximated by  
$$
\frac{V}{L^2} \sum_{l} l^2 (\sum_{p, l_p = l}
\frac{A_p}{r_p})^2$$
 where the first sum is over isometry classes of periodic orbits of
 period $2n$  and the second is over a given isometry class.

A combinatorial analysis of periodic orbits leads  (\cite{Ber})
to a formula valid  near $\tau =0$, yielding a small $\tau$ expansion 
 $K_{st}(\tau) = 1-4\tau + 8\tau^2 -\frac{8}{3}\tau^3 +
o(\tau^4)$. 
\\One should note that the first three terms are
given by the orbits consisting of only one edge; to compute the
$\tau^3$ and $\tau^4$ terms, one also needs to take into account 
the isometry classes of orbits 
consisting of two or three different edges, 
but one can then only take into account in these isometry classes 
the orbits
with the maximum number
of backscattering (that is, orbits of the form $aa...abb...b$, or
$aa...abb...bcc...c$). This approach starts only to fail for the $\tau^5$
term, whereas the diagonal approximation fails at $\tau^4$ (\cite{ThB}).

\section{The form factor for the quasar graphs if $\nu_3=0$}


We thus want to compute for 
$\tau \in[ \frac{n}{V}
-\frac{1}{2V},\frac{n}{V}+\frac{1}{2V}] $ fixed, 
$$K(\tau) = \frac{V}{4V^2}\sum_{{\cal{L}}}l^2 (\sum_{p\in
  {\cal{L}}} \frac{A_p}{r_p})^2 \simeq  \tau^2 V \sum_{{\cal{L}}} (\sum_{p\in
  {\cal{L}}} \frac{A_p}{r_p})^2$$ 

Some terms in the sum over equivalence classes concern periodic orbits
 visiting only one star graph (and no glueing edge). 
This yields two subsums very similar to the form factor of one star graph :
the difference is that in the subsum over the $i$-th stargraph, 
the number of available vertices is $V_i$, whereas
the scattering and backscattering factors are $t_i=\frac{2}{V_i+M}$ and $r_i=(-1+\frac{2}{V_i+M})$, and not $\frac{2}{V_i}$ and $(-1+\frac{2}{V_i})$.
 Nevertheless, if $M=o(V)$, these two sums become in the limit $(n,V)$ tends to infinity $\nu_1 K_{st}(\frac{\tau}{\nu_1}) + \nu_2 K_{st}(\frac{\tau}{\nu_2 })$, where $K_{st}$ is the form factor for a star graph presented in  the preceding section.

We will compute the corrections of the most backscattering orbits to these 
unglued terms, and check that they disappear in the limit. Since these orbits should give the correct $\tau\rightarrow 0$ asymptotics, this means that the
form factor of the quasar graph is in this case  $\nu_1 K_{st}(\frac{\tau}{\nu_1}) + \nu_2 K_{st}(\frac{\tau}{\nu_2 })$, and that the glueing has no effect.
In particular, if $\nu_1 =\nu_2 = \frac{1}{2}$, one gets $K_{st}(2\tau)$.

Let us  consider orbits of length around $2n$ with maximal
backscattering, that is orbits consisting only of one glueing edge.
Each equivalence class consists of one orbit $\alpha...\alpha$ with repetition number $n$,
and $A_p= (r_1 r_2)^n$. There are $M$ such equivalence classes. Hence their
 contribution to $K$ is $\frac{1}{4V} M (2n)^2
\frac{(r_1 r_2)^{n}}{n^2}$ which, when $V,n$ tend to infinity such that
$\frac{n}{V}$ is finite, is of the order $\frac{M}{V}$, hence
negligible if $M=o(V)$. This is not surprising : there are not enough
such terms for them to contribute. This will be the case for all orbits
containing glueing edge : to make up for the $\frac{1}{V}$ terms corresponding to normal scatterings, one needs a $V$ combinatorial factor, that one does not
get when using glueing edges.


\section{The $\nu_3 \neq 0$ case.}

The situation is very different here, since even the two subsums do not
give a proper star graph limit. We will thus here compute the exact contributions of the most backscattering orbits of length around $2n$. 

\subsection{Orbits with no scattering}

Let us first examine the case of the no scattering orbits. They consist of one edge, and they can be of three kinds, depending if the edge is a glueing edge or not. In all cases, the equivalence class is a singleton.

If the edge is on star graph $i$, one gets $A_p=r_i^n$ and a
repetition number $n$. There are $V_i$ such equivalence classes, yielding
in the limit 
$\nu_i \exp(-4\frac{\tau}{\nu_i+\nu_3})= \nu_i \exp(-4\tau_i)$ if one writes
$\tau_i = \frac{\tau}{\nu_i+\nu_3}$.

If the edge is a glueing edge, $A_p = (r_1r_2)^n$ (remember that there is no
trivial backscattering for glueing edges). There are $M$ such classes, giving a total contribution of
$\nu_3 \exp(-4\tau (\frac{1}{\nu_1+\nu_3} + \frac{1}{\nu_2+\nu_3})) = \nu_3 \exp(-4(\tau_1+\tau_2))$.

\subsection{Orbits with two scatterings}

These are orbits with two different edges. There are three different cases. 

If these two edges belong to the same star graph $i$, the weight $A_p$ 
is 
$t_i^2r_i^{n-2}$.
There are $\frac{V_i(V_i-1)}{2}$ ways to choose the edges, and $n-1$ ways to choose how many times each 
orbit is visited.

This yields when $V$ tends to infinity a contribution to the form factor  $8\tau^3 \frac{\nu_i^2}{(\nu_i+\nu_3)^4} \exp(-4\tau_i)= 8\tau^3 \frac{\nu_i^2}{(\nu_i+\nu_3)^4} + o(\tau^3)$.

If the two edges are glueing edges, there are $M(M-1)/2$ ways to choose them.  But the weights  depend on the parity of the number
of visits of each edge. If it is odd, there is one scattering on star graph 1 
and one scattering on star graph 2, hence a weight 
$A_p = t_1t_2(r_1r_2)^{n-1}$.
If it is even, the scatterings can happen either on star graph 1 or on star graph 2 : there are two orbits giving the main contribution in the equivalence class, with respective weights 
$A_p = t_1^2 r_1^{n-2} r_2^n$ and $A_p = t_2^2 r_2^{n-2} r_1^n$.
Even or odd, there are $n$ or $n-1$ ways to choose how many times each orbit is visited. 

All in all, this case yields when $V$ tends to infinity a contribution to the form factor  of
$8\tau^3 \nu_3^2 \exp(-4\tau_2) \exp(-4\tau_1)(\frac{1}{(\nu_1+\nu_3)^2(\nu_2+\nu_3)^2} + (\frac{1}{(\nu_1+\nu_3)^2} + \frac{1}{(\nu_2+\nu_3)^2})^2)
=8\tau^3 \nu_3^2 (\frac{1}{(\nu_1+\nu_3)^2(\nu_2+\nu_3)^2} + (\frac{1}{(\nu_1+\nu_3)^2} + \frac{1}{(\nu_2+\nu_3)^2})^2) +o(\tau^3)$

Last, if one of the two edges is a glueing edge and the other one is a $i$-star graph edge, there are $MV_i$ ways to choose the edges; there are $n-1$ ways to choose 
the number of visits of each edge, but the weight depends on it : for example,
 if $i=1$, and if the glueing edge is visited $n_e$ times ($n_e$ has to be even), the weight is 
$A_p = t_1^2 r_1^{n-2} r_2^{\frac{n_e}{2}}$. The sum over equivalence classes
involves a sum over $n_e$ which is a geometric sum; in the limit $V\rightarrow \infty$ one gets a term
$$\tau^2 \frac{4\nu_3\nu_1(\nu_{2}+\nu_3)}{(\nu_1 + \nu_3)^4} \exp(-4\tau_1) (1-\exp(-4\tau_2)) =\tau^3 \frac{16\nu_3\nu_1}{(\nu_1 + \nu_3)^4} +o(\tau^3). $$

All in all, the first contribution of the two scatterings orbits to the 
small $\tau$ expansion of the form factor is a $\tau^3$ term equal to 
$$8\tau^3\left(\frac{1}{(\nu_1+\nu_3)^2} + \frac{1}{(\nu_2+\nu_3)^2} + \frac{3\nu_3^2}{(\nu_1+\nu_3)^2(\nu_2+\nu_3)^2}\right).$$

\subsection{$\tau^3$ expansion for the $\nu_3\neq 0$ case}
Putting everything together, the $\tau^3$ expansion of form factor for 
the quasar graph is
$$1-8\tau+8\,{\frac {1+3\nu_3}{ \left( {\it \nu_2}+{\it \nu_3} \right)  \left( {\it\nu_1}+{\nu_3}
 \right) }}{\tau}^{2}-\frac{8}{3}\,{\frac {{{\it \nu_1}}^{2}+14\,{\it \nu_3}\,{\it \nu_1}+
14\,{\it \nu_3}\,{\it \nu_2}+17\,{{\it \nu_3}}^{2}+{{\it \nu_2}}^{2}}{ \left( {
\it \nu_2}+{\it \nu_3} \right) ^{2} \left( {\it \nu_1}+{\it \nu_3} \right) ^{2}}}{
\tau}^{3}+o \left( {\tau}^{3} \right) .$$

\begin{figure}
\centering\epsfig{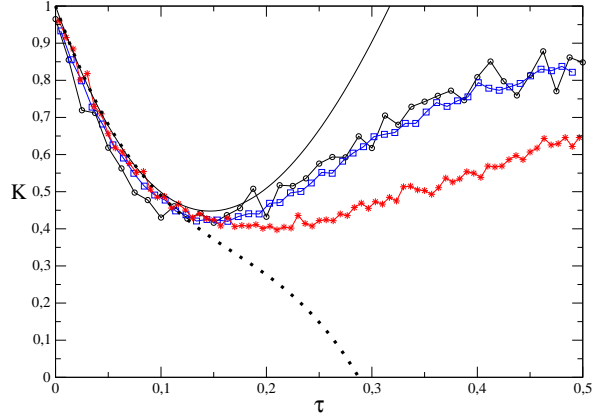}
\caption{$K^*(2\tau)$ (circles), and the form factors for a $(50,50,1)$ quasar graph (squares) and for a $(50,50,50)$ quasar graph (stars). The solid line
corresponds to the $\tau^3$ expansion in the $\nu_3=0$, $\nu_1=\nu_2$ case, the dashed line
corresponds to the $\tau^3$ expansion in the $\nu_1=\nu_2=\nu_3$ case. }
\label{Factnum} 
\end{figure}

In the particular case $\nu_1 = \nu_2$, (and hence $\nu_3 = 1-2\nu_1$), one gets
$$ 1-8\tau +\frac{16(1-3\nu_1)}{(1-\nu_1)^2} \tau^2 - \frac{8}{3}\frac{14\nu_1^2 - 40\nu_1 + 17}{(1-\nu_1)^4} \tau^3 + o(\tau^3).$$
The $\tau^2$ coefficient is then comprised between $32$ (corresponding to $\nu_3=0$) and $36$ (corresponding to $\nu_1=\nu_2=\nu_3)$. The difference of the 
$\tau^3$ term with the unglued $\nu_3=0$ case is also maximal for $\nu_1=\nu_2+\nu_3$. Yet, even in this case, the two expansions in the glued case and the
unglued case are very similar for $\tau\in[0,1]$, so that it is 
difficult to give numerical evidence. As shown on figure \ref{Factnum}, the
two form factors are really close to each other until $\tau \simeq 0.15$, and
one would then need higher order terms.

\section{Conclusion}
The first effect of the glueing is in the $\tau^2$ term : the non-universal 
statistics of the quantum graph seems to be quite robust. This is not so
surprising, since this non universal statistics is strongly linked to the fact that the random walk
on the graph converges very slowly to the uniform distribution, and this fact
cannot be changed by the glueing. It should be expected that
any ``star-graph'' component of a quantum graph, as long as it is macroscopically big (containing a non-zero fraction of the vertices), should yield
a form factor such that $K(0)=1$.

This work benefitted from the support of the french ANR project ``Arithmatrix''.
\section*{References}
\bibliographystyle{unsrt}
\bibliography{qgraph}
\end{document}